\documentclass[aip,cha,amsmath,amssymb,reprint]{revtex4-1}
\usepackage{xcolor,cancel}
\usepackage{graphicx}
\usepackage{soul}
\graphicspath{ {figures/} }
\usepackage{bbm}
\usepackage{subcaption}

\newcommand{\I}{\ensuremath{\mathbbm{i}}}

\renewcommand{\L}{\mathcal{L}}

\draft 

\begin{document}


\title{Instanton based importance sampling for rare events in stochastic PDEs} 



\author{Lasse Ebener}
\affiliation{Institut f\"ur Theoretische Physik I, Ruhr-Universit\"at Bochum, Universit\"{a}tsstra\ss{}e 150, 44780 Bochum, Germany}
\author{Georgios Margazoglou}
\affiliation{Dept. Physics and INFN, University of Rome ``Tor Vergata'',
Via della Ricerca Scientifica 1, I-00133 Roma, Italy}
\affiliation{Computation-based Science and Technology Research Center, Cyprus Institute, 20 Kavafi Str., 2121 Nicosia, Cyprus}
\author{Jan Friedrich}
\affiliation{Institut f\"ur Theoretische Physik I, Ruhr-Universit\"at Bochum, Universit\"{a}tsstra\ss{}e 150, 44780 Bochum, Germany}
\affiliation{Laboratoire de Physique, Ecole Normale Sup\'erieure de Lyon,
CNRS, Universit\'e de Lyon, F-69007, Lyon, France}
\author{ Luca Biferale}
\affiliation{Dept. Physics and INFN, University of Rome ``Tor Vergata'',
Via della Ricerca Scientifica 1, I-00133 Roma, Italy}
\author{Rainer Grauer}
\affiliation{Institut f\"ur Theoretische Physik I, Ruhr-Universit\"at Bochum, Universit\"{a}tsstra\ss{}e 150, 44780 Bochum, Germany}

\def\NOTE#1{{{\bf [#1]}}}
\newcommand*\mystrut[1]{\vrule width0pt height0pt depth#1\relax}
\date{\today}

\begin{abstract}
We present a new method for sampling rare and large fluctuations in a
non-equilibrium system governed by a stochastic partial differential equation
(SPDE) with additive forcing. To this end, we deploy the so-called instanton
formalism that corresponds to a saddle-point approximation of the action in the
path integral formulation of the underlying SPDE.  The crucial step in our
approach is the formulation of an alternative SPDE that incorporates knowledge
of the instanton solution such that we are able to constrain the dynamical evolutions around extreme flow configurations only. Finally, a reweighting procedure based on the Girsanov theorem is applied to recover the full distribution function of
the original system. The entire procedure is demonstrated on the example of the
one-dimensional Burgers equation. Furthermore, we compare our method to conventional direct numerical simulations
 as well as to Hybrid Monte Carlo methods. It will be shown
that the instanton-based sampling method outperforms both approaches and allows for an accurate quantification
of the whole probability density function of velocity gradients from the core to the very far tails.
\end{abstract}
\maketitle

\section{Introduction} \label{sec:intro}

Non-equilibrium systems that possess a large number of interacting degrees of
freedom typically exhibit strongly anomalous statistical properties which can be
attributed to rare large fluctuations. Typical examples include the occurrence
rate of earthquakes~\cite{knopoff1977}, the existence of rogue
waves~\cite{onorato:2013,dematteis:2018,hadjihosseini:2018}, crashes in the
stock market~\cite{Ghashghaie1996,longin2016,bouchaud-potters:2009}, the
occurrence of epileptic seizures~\cite{Lehnertz2006}, or perhaps the most
enigmatic case, the distribution of velocity increments in hydrodynamic
turbulence~\cite{Frisch:1995}. In turbulence theory, a central notion is
the energy cascade which implies a non-linear and chaotic transfer between different scales \cite{alexakisbiferale2018}.
Although well-established descriptions by Richardson, Kolmogorov, Onsager, Heisenberg, and others (see the reviews~\cite{Frisch:1995,monin}) can capture the mean field features of the cascade
process in a phenomenological way, the nature of small-scale turbulent energy
dissipation is far less understood and is usually attributed to nearly singular
localized vortical structures~\cite{yeung:2015,debue:2018}. Empirically, the
energy transfer from large to small scales is accompanied by a breaking of
self-similarity of the probability density function (PDF) of  velocity
increments, a phenomenon called {\it intermittency}.  Intermittency is intimately connected to non-Gaussian statistics and extreme events and is often described in a statistical sense, using random multiplicative cascades leading to multifractal measures \cite{Frisch:1995,benzibiferale2009}. On the other hand, it manifests itself by the presence of singular or quasi-singular structures, highly concentrated in a few spatial locations. From a mathematical point of view, large fluctuations of the fluid variables are controlled by the theory of large deviations~\cite{freidlin1998random,varadhan2016,Touchette2009,cecconi:2014},
which is concerned with the exponential decay of the PDF for large field values, see, e.g., the paper\citep{Bouchet2014} for a  numerical application  based on the Onsager-Machlup functional in the context of geophysical flows.

Besides numerical large deviation methods, cloning and selection
strategies (that favor the desired event) have evolved into mature simulation techniques, too~\cite{grassberger2002,Giardina2011}. Recently, such methods have been deployed in order to
investigate return times in Ornstein-Uhlenbeck processes, in the
drag forces acting on an object in turbulent flows~\cite{lestang:2018}, as well
as for extreme heat waves in weather dynamics~\cite{ragone2018}. Other
numerical methods include  direct importance sampling in configuration
space~\cite{dellago1998}, a modification of  transition path
methods~\cite{Touchette2009} in form of the so-called string
method~\cite{weinan2002} and the geometric minimum action method\cite{heymann-eijnden:2008}.

In this paper we are interested to apply saddle point techniques to estimate extreme events (also called instantons or optimal fluctuations) as originally introduced in the context of solid state disordered systems \cite{lifshitz:1964,halperin-lax:1966,zittartz-langer:1966,langer:1967,langer:1969}
(see also \cite{grafke-grauer-schaefer:2015} for an overview). Especially, we refer to the
works of Zittartz and Langer \cite{zittartz-langer:1966,langer:1967,langer:1969}
which contain nearly all the recipes we are using today.
Single- and multi-instantons dynamics have often been advocated as some of the possible mechanisms of anomalous fluctuations in hydrodynamical systems and models thereof ~\cite{Balkovsky:1996cc,Balkovsky:1997zz,depietro2017,mailybaev2013,daumont2000}.
We will use the so-called
Janssen-de Dominicis~\cite{janssen:1976,dedominicis:1976} path integral
formulation of the Martin-Siggia-Rose (MSR) operator technique
~\cite{martin-siggia-rose:1973} for classical stochastic systems.
In particular, we will apply it to the important case of the one-dimensional stochastically forced  Burgers equation:
\begin{equation}
	u_t(x,t) + u(x,t)u_x(x,t)	= \nu u_{xx}(x,t) + \eta(x,t)\;,
	\label{eq:burgers}
\end{equation}
 where the nonlinearity tends to form shock
fronts that are ultimately smeared out by viscosity and lead to the appearance
of large negative velocity gradients (see below for further details on the equations). The Burgers equation constitutes a
high-dimensional and highly non-trivial example of a complex system with fluctuations far away from Gaussianity. The Burgers equation can be also considered as
a simplified one-dimensional compressible version of the Navier-Stokes equation and has been extensively studied in the past
decades~\cite{Gotoh:1994,Chekhlov:1995a,Chekhlov:1995b,Bouchaud:1995zza,Gurarie:1995qc,Balkovsky:1996cc,Balkovsky:1997zz,Falkovich:1995fa,Polyakov:1995mn,Boldyrev:1996eu,E:2000,e-eijnden:2000a,e-eijnden:2000b,friedrich:2018}
(see also the review~\cite{Bec:2007}) using
numerical simulations~\cite{Chekhlov:1995a,Chekhlov:1995b,Gotoh:1994}, the
replica method~\cite{Bouchaud:1995zza}, operator product
expansion~\cite{Polyakov:1995mn,Boldyrev:1996eu,Lassig:1998hb,friedrich:2018}, asymptotic methods~\cite{E:2000,e-eijnden:2000a,e-eijnden:2000b} as
well as instanton
methods~\cite{Gurarie:1995qc,Balkovsky:1996cc,Balkovsky:1997zz,Falkovich:1995fa,grafke-grauer-schaefer:2015}.
Similarly, the Kardar-Parisi-Zhang equation, which is strictly connected to (\ref{eq:burgers}), has also recently been studied using instantons~ \cite{janas-kamenev-meerson:2016,smith-kamenev-meerson:2018}.\\

The main problem when dealing with instanton approximations of the whole probability distribution is to evaluate the fluctuations around them, which is, in turn,  connected to the most important problem of quantifying the influence of the saddle-point solution to all field values, including the ones that are not extremal.

In this paper, we propose a new method to study the shape of the PDF of the Burgers velocity gradients $u_x$,  in those parameter regions where it is dominated by instantons, considering also the fluctuations around the saddle point configurations. To do that, we will decompose the velocity field in the
MSR action into a contribution that stems from the instanton as well as a
fluctuation around this object. We then proceed and derive an evolution equation
for the fluctuation in the {\it background} of the  instanton solution for a given gradient.
A subsequent reweighting procedure allows us to calculate the full PDF with a
much more accurate description of the tails in
comparison to ordinary direct numerical simulations (DNS) of Burgers turbulence. We also show that our method is  computationally less substantially challenging than other approaches based on Markov Chain Monte Carlo methods to generate extreme and rare flow configurations \cite{Margazoglou2018}. Hence, the method can be
considered as an optimal application of rare events importance sampling and we call it the {\it Instanton based Importance Sampling} (IbIS), see also the work \cite{buhler2007} for a similar idea. In our formulation, the method is general enough to be applied to many different SPDEs.

The paper is organized as follows: In section \ref{sec:path_integral_instantons}
we review the path integral formulation of stochastic systems. Section
\ref{sec:reweighting} constitutes the core of the paper, where we present our reweighting
procedure. In section \ref{sec:num_sim_results} we  describe in detail the
numerical protocol and we compare the results obtained with IbIS against those obtained using DNS and a Hybrid Monte Carlo approach \cite{Margazoglou2018}. We close with a
summary and an outlook on further applications.

\section{Path integral formulation and instantons}  \label{sec:path_integral_instantons}

To make our exposition self-consistent, here we describe the path integral
formulation of stochastic systems, the subsequent derivation of the instanton
equations, and the calculation of fluctuations around the instanton using an
appropriate reweighting procedure. The presentation follows closely the seminal
work of \citet{Balkovsky:1996cc}.

\subsection{Path integral formulation of stochastic systems}

The Martin-Siggia-Rose-Janssen-de Dominicis formalism (hereinafter referred to as
MSRJD formalism) \cite{martin-siggia-rose:1973,janssen:1976,dedominicis:1976}
was developed in the early 1970's to calculate statistical properties of
classical systems using a path integral formulation. Following the same notation as
in\cite{chernykh-stepanov:2001,grafke-grauer-schaefer:2015}, we consider a
stochastic differential equation
\begin{equation}\label{eq:sde}
\dot{u} + N[u,x] = \eta (x,t)\	,
\end{equation}
where $\eta$ is an additive Gaussian noise with correlation
\begin{equation}\label{eq:noise_corr}
\langle \eta (x,t) \eta (x+r,t+s) \rangle = \chi (r) \delta (s)\ .
\end{equation}
Here, the $\delta$-correlation implies that the forcing $\eta$ is white in time,
while $\chi (r)$ is some arbitrary spatial correlation. Considering that from
Eq.~(\ref{eq:sde}) the field $u$ is a functional $u[\eta]$ of the forcing
$\eta$, we introduce the MSRJD formalism as follows. The expectation value of an
observable $\left< \mathcal{O}[u] \right>$, as the average over all possible
noise realizations, can be defined as
\begin{equation}
\langle \mathcal{O}[u] \rangle = \int \mathcal{D}\eta\, \mathcal{O}[u[\eta]]\, e^{- \int dt \,\langle \eta, \chi^{-1} \eta \rangle /2}\	,	\label{eq:path_int}
\end{equation}
where the integral in the exponent derives from $\eta$ being normally
distributed, with $\langle\ \cdot\ ,\ \cdot\ \rangle$ being the $L^2$ inner
product. Changing the integration from $\eta$ to $u$, given Eq.~(\ref{eq:sde}),
modifies the measure in Eq.~(\ref{eq:path_int}) as $\mathcal{D}\eta = J[u]\,
\mathcal{D}u$, where $J[u] = \mathrm{det} \left( \frac{\delta\eta}{\delta u}
\right) = \mathrm{det} \left( \partial_{t} + \frac{\delta N}{\delta u} \right)$
is the Jacobian associated to the map $\eta\mapsto u$. This results into what is
called the Onsager-Machlup functional\cite{Onsager:1953}
\begin{equation}
\langle \mathcal{O}[u] \rangle = \int \mathcal{D}u\, \mathcal{O}[u] J[u]\, e^{- \int dt \, \langle \dot{u} + N[u], \chi^{-1} (\dot{u} + N[u]) \rangle /2}\	.
\end{equation}
It is the starting point for direct minimization of the Lagrangian action
\begin{equation} S_{\mathcal{L}}[u, \dot{u}] = \frac{1}{2} \int dt \, \langle \dot{u}
+ N[u], \chi^{-1} (\dot{u} + N[u]) \rangle \ . \label{eq:org_action}
\end{equation}

Most of the time, it is more convenient to work with the original correlation
function $\chi$ instead of its inverse. We therefore perform a
Hubbard-Stratonovich transformation, which introduces an auxiliary field
$\tilde{p}$ and by virtue of a Fourier transform and completing the square
eliminates the inverse of the correlation function $\chi^{-1}$ and in addition
the noise appears only linearly in the action:
\begin{equation}
\langle \mathcal{O}[u] \rangle = \int \mathcal{D}\eta \mathcal{D}\tilde{p}\, \mathcal{O}[u[\eta]]\,
e^{- \int dt \, [ \langle \tilde{p} , \chi \tilde{p} \rangle /2 - \I \langle \tilde{p} , \eta \rangle ]}\	.	\label{eq:aux_path_int}
\end{equation}

We then execute the coordinate transformation from $\eta$ to $u$
\begin{equation}
\langle \mathcal{O}[u] \rangle = \int \mathcal{D}u \mathcal{D}\tilde{p}\,
\mathcal{O}[u] J[u]\,
e^{- S[u,\tilde{p}]},
\end{equation}
with the action function $S[u,\tilde{p}]$ given by
\begin{equation*}
S[u,\tilde{p}] = \int dt \, \left[ -\I \langle \tilde{p} , \dot{u} + N[u] \rangle
+ \frac{1}{2} \langle \tilde{p} , \chi \tilde{p} \rangle \right] \	.
\end{equation*}
For the purpose of convenience to obtain an Euclidian path integral we now
substitute $\tilde{p}$ with $p$, by use of the relation $\tilde{p} = \I\, p$ and
obtain the Hamiltonian action
\begin{equation}
S_{\mathcal{H}}[u,p] = \int dt \, \left[ \langle p , \dot{u} + N[u] \rangle
- \frac{1}{2} \langle p , \chi p \rangle \right] \	.	\label{eq:s_u_p}
\end{equation}

The next step is to minimize this action functional to obtain the instanton solutions.

\subsection{Instantons}\label{subsec:instantons}

We are interested in rare and large fluctuations of velocity gradients
$u_x(x,t)$ in the Burgers equation (\ref{eq:burgers}). Since this chaotic and
turbulent system is invariant under time translations and Galilean
transformations, the probability function of velocity gradients can be cast into
the following form as a path integral
\begin{align}\label{eq:vel_grad_pi}
\mathcal{P}(a) &= \langle \delta( u_{x}(0,0) - a ) \rangle \nonumber \\
&= \int \mathcal{D}u\, \mathcal{D}p\, \int_{-\I \infty}^{\I \infty} d\mathcal{F}\, \mathrm{exp}\{ - S_{\mathcal{H}} + \mathcal{F}\, [ u_{x}(0,0) - a ] \}\ .
\end{align}
Here $\mathcal{F}$ stems from the Fourier transformat of the
$\delta$-function and serves as a Lagrange multiplier. Hence, from
Eq.~(\ref{eq:s_u_p}) we have
\begin{align}\label{eq:aux_action}
S_{\mathcal{H}}[u,p] =\ &\int_{- \infty}^{0} dt\, \int dx\, p(x,t)\,
( u_{t} + u u_{x} - \nu u_{xx} ) \nonumber \\
&-\frac{1}{2} \int_{- \infty}^{0} dt\, \int dx\, dx'\, p(x,t)
\chi(x - x') p(x',t)\ .
\end{align}

Now $\mathcal{F}$ is treated as a large parameter so that the saddle point
approximation can be used in order to derive instanton configurations, i.e.
``classical'' solutions that minimize the action and therefore dominate the path
integral of Eq.~(\ref{eq:vel_grad_pi}). The instanton equations are obtained
from the conditions
\begin{equation}
\frac{\delta S}{\delta u} = 0 \quad \text{and}\quad \frac{\delta S}{\delta p} = 0\ .
\end{equation}
When carried out, the functional derivatives above yield the so called instanton equations:
\begin{align}
&u_{t} + u u_{x} - \nu u_{xx} = \chi*p\, , \label{eq:u_inst} \\
&p_{t} + u p_{x} + \nu p_{xx} = \mathcal{F}\, \delta(t)\, \delta'(x)\, \label{eq:p_inst} ,
\end{align}
where $u(x,t)$ and $p(x,t)$ have the following boundary conditions:
\begin{align*}
	&\ \underset{t\rightarrow -\infty}{\mathrm{lim}}\ u(x,t) = 0 \qquad \ \
	\underset{t\rightarrow -\infty}{\mathrm{lim}}\ p(x,t) = 0
	\\
	&\underset{|x|\rightarrow +\infty}{\mathrm{lim}}\ u(x,t) = 0 \qquad
	\underset{|x|\rightarrow +\infty}{\mathrm{lim}}\ p(x,t) = 0
\end{align*}
and $\chi*p$ is the convolution
\begin{equation}
(\chi*p)(x) = \int dx'\, \chi(x-x')\, p(x',t)\ .
\end{equation}
Because of the $\delta$-function, the RHS of Eq.~(\ref{eq:p_inst}) is an
initial condition for $p$.
Furthermore, the RHS of Eq.~(\ref{eq:u_inst}) will often be abbreviated:
\begin{equation}
	\chi*p = P\	.
\end{equation}
Making use of Eqs.~(\ref{eq:aux_action}) and (\ref{eq:u_inst}) we may calculate
the instanton action
\begin{equation}
S^{I(a)} = \frac{1}{2}\, \int_{-\infty}^{0} dt\, \int dx\, dx'\, p(x,t)\, \chi(x - x')\, p(x',t)\ ,	\label{eq:inst_action}
\end{equation}
where $I(a)$ denotes that the instanton has a gradient of $u_x = a$. We denote all
other instanton related quantities in a similar way.

\section{Instanton based Reweighting}\label{sec:reweighting}

The process of reweighting allows us to assemble the PDF related to the
stochastic Burgers equation by solving the stochastic PDE for the fluctuations
around the instanton.
In order to derive the instanton equations, we considered the minimum of the
Janssen-de Dominicis action $S_{\mathcal{H}}[u,p]$. However, to
derive this stochastic PDE, we will work again with the original
Onsager-Machlup action (see Eq.~(\ref{eq:org_action})):
\begin{align}
	S_{\mathcal{L}}[u,\dot{u}] &= \frac{1}{2} \int dt \, \langle \dot{u} + N[u], \chi^{-1} (\dot{u} + N[u]) \rangle  \nonumber
	\\
	&= \frac{1}{2} \int dt\, dx\, dx' \, \left[u_{t} + u u_{x} - \nu u_{xx} \right]  \\
	 & \qquad \times \, \chi^{-1}(x-x')\, \left[u_{t} + u u_{x'}+ \nu u_{x'x'} \right].\nonumber
\end{align}
We then decompose the field into instanton and fluctuation
\begin{equation}
	u = u^{I(a)} + \delta u\	.
\end{equation}
This results in
\begin{align}
	S_{\mathcal{L}} &= \frac{1}{2} \int dt\, dx\, dx'\,
	\Big[P^{I(a)} + \delta u_{t} + \delta u \delta u_{x} - \nu \delta u_{xx} \nonumber \\
	& \qquad+ (u^{I(a)} \delta u)_{x} \Big]\,
	\chi^{-1}(x-x')\, [\mathellipsis]	\nonumber
	\\
	&= S^{I(a)} + \tilde{S}^{a} - \frac{1}{2} \int dt\, dx\, p^{I(a)}_{x} (\delta u)^{2}\	,
	\label{eq:action_expansion}
\end{align}
with
\begin{align}
	\tilde{S}^{a} = \frac{1}{2} \int dt\, dx\, dx'\,&
	[\delta u_{t} + \delta u \delta u_{x} - \nu \delta u_{xx} + (u^{I(a)} \delta u)_{x}]\, \nonumber\\
	& \qquad \times \chi^{-1}(x-x')\, [\mathellipsis]\	.	\label{eq:tilde_action}
\end{align}
Here $[\mathellipsis]$ denotes the appropriate expression evaluated at the
position $x'$. Note that all linear variations vanish by definition
of the instanton. Thus we define
\begin{equation}
	\Delta S^a = S_{\mathcal{L}} - \tilde{S}^{a} = S^{I(a)} - \frac{1}{2} \int dt\, dx\, p^{I(a)}_{x} (\delta u)^{2}\	.	\label{eq:delta_s}
\end{equation}
Now to derive the stochastic equation corresponding to the action
$\tilde{S}^{a}$, we reverse the derivation of the path integral formulation and
obtain
\begin{equation}
	\delta u_{t} + \delta u \delta u_{x} - \nu \delta u_{xx} = \eta - (u^{I(a)} \delta u)_{x}\	.	\label{eq:tilde_pde}
\end{equation}

Next, we have to change the path measure for this process in order to connect
the statistics to the original one. To do that, we first  consider the identity
\begin{align}
	P_{S_{\mathcal{L}}}(s) &:= \delta(u_x(0,0)-s) e^{-S_{\mathcal{L}}} \nonumber\\
	&= \delta(u_x(0,0)-s) e^{-(\tilde{S}^a + \Delta S^a)} \nonumber\\
	&= \delta(\delta u_x(0,0)+a-s) e^{-\tilde{S}^a} e^{-\Delta S^a},
   \label{eq:key}
\end{align}
where $P_{S_{\mathcal{L}}}(s)$ denotes the  path measure of the distribution of gradients, $s=u_x(x,t)$ at $(x,t)=(0,0)$  in the original Burgers equation as one would obtain it  by performing numerical simulations of
(\ref{eq:burgers}).  On the other hand,
if we sample events for the gradient of the fluctuations, $\delta u$, around the instanton $I^{(a)}$, i.e. when  $u_x = a$ through the new stochastic
equation~(\ref{eq:tilde_pde}) we would get a PDF generated by the measure $e^{-\tilde{S}^a}$. The last equality in (\ref{eq:key}) tell us that in order to get the unbiased original PDF we need to reweight with a factor $e^{-\Delta{S}^a}$.
 A similar approach was formulated in a simpler setting by \citet{buhler2007}.

\begin{figure}[h!tbp]
	\centering
	\includegraphics[width=0.5\textwidth]{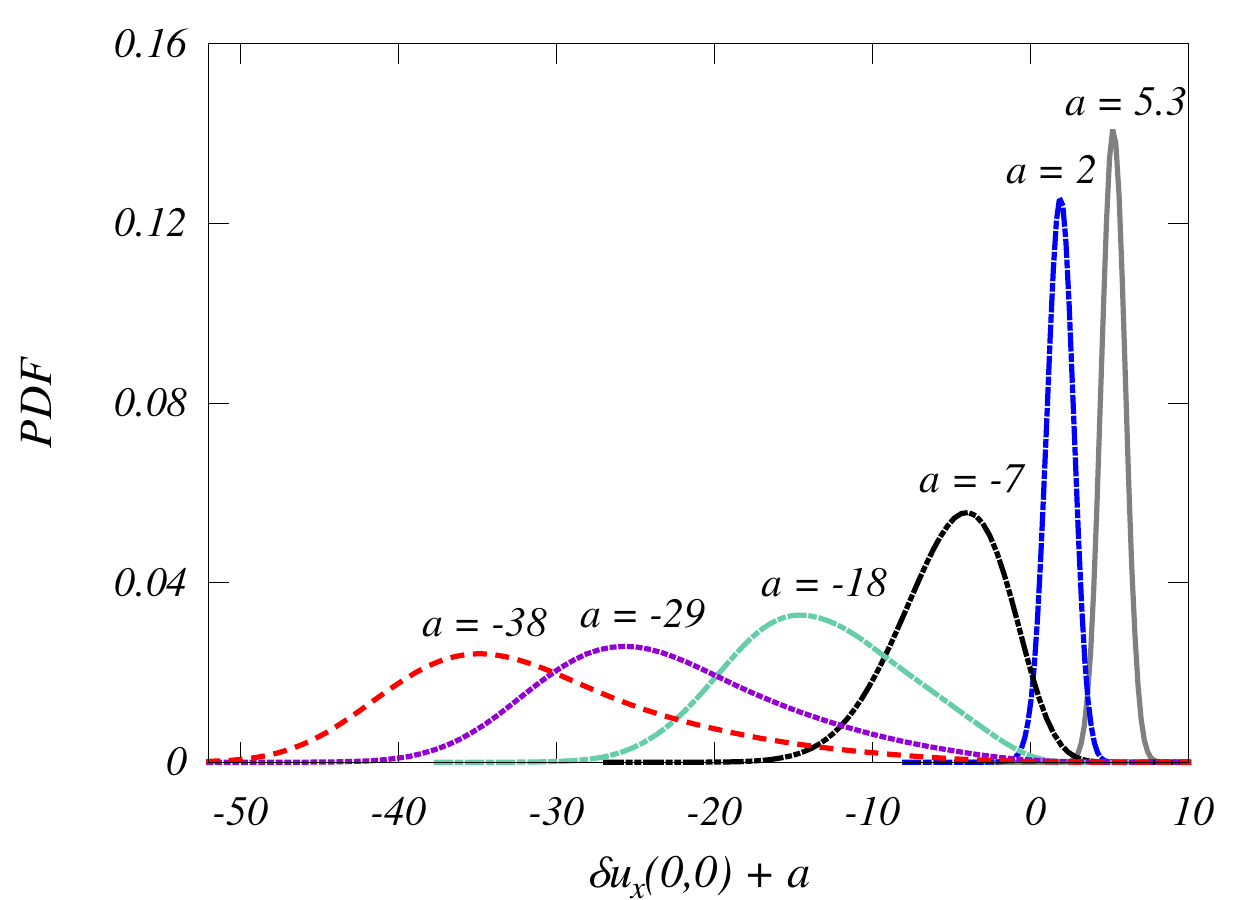}
	\caption{Shape of the PDFs for the gradient of the fluctuations $\delta u_x$ (shifted by $a$)
            around the instanton $u^{I(a)}$ for six different values of $u_x=a$.}
	\label{fig:gaussian_around_instanton}
\end{figure}

If we would proceed in choosing one value $u_x=a$ to calculate the instanton
with $u_x^{I(a)}(x=0,t=0) = a$, we will be able to sample the statistics near this
value very efficiently. However, values $u_x = s$ far away from $a$ will be
sampled with poor performance. Thus a major step is to choose
$s=a$, which means that for every point $u_x =a$ in the PDF we first calculate
the instanton and then using Eq.~(\ref{eq:delta_s}) obtain the PDF at
$u_x = a$. This procedure is further motivated in Fig.~\ref{fig:gaussian_around_instanton}.
This figure shows the PDFs for the gradient of the fluctuations $\delta u_x (0,0)$ around the instanton $u^{I(a)}$, measured at (0,0) and shifted by $a$ (for comparison) using six different values of $u_x=a$, obtained from simulations of Eq.~(\ref{eq:tilde_pde}).
Thus the actual form of the Girsanov transformation used in our instanton reweighting approach reads
\begin{align}
	P_{S_{\mathcal{L}}}(s) &:= \delta(u_x(0,0)-s) e^{-S_{\mathcal{L}}} \nonumber \\
	&= \delta(\delta u_x(0,0)) e^{-(\tilde{S}^a + \Delta S^a)}  \label{eq:P_tildeS}
\end{align}

\section{Numerical simulations of rare events} \label{sec:num_sim_results}

In this section we describe the numerical procedure to calculate the full PDF of
velocity gradients step by step. The numerical integration of the stochastic PDE
given in Eq.~(\ref{eq:tilde_pde})
\begin{equation*}
\delta u_{t} + \delta u \delta u_{x} - \nu \delta u_{xx} = \eta - (u^{I(a)} \delta u)_{x}\
\end{equation*}
is achieved using the Euler-Maruyama method \cite{Higham2001} in combination
with an integrating factor \cite{Canuto:1988} for the dissipative term. The
spatial correlation function of the forcing (\ref{eq:noise_corr}) follows a
power law proportional to $k^{-3}$ in Fourier space and has a cutoff at $k_F =
N_x/3$, where $N_x$ is the spatial number of grid points. The nonlinear term is
evaluated with the pseudospectral method.  The instanton equations
(\ref{eq:u_inst})-(\ref{eq:p_inst}) are solved using an iterative method as
described in Chernykh and Stepanov
\cite{chernykh-stepanov:2001,grafke-grauer-schaefer:2015}.

\begin{figure}[t!bp]
	\centering
	\includegraphics[width=0.5\textwidth]{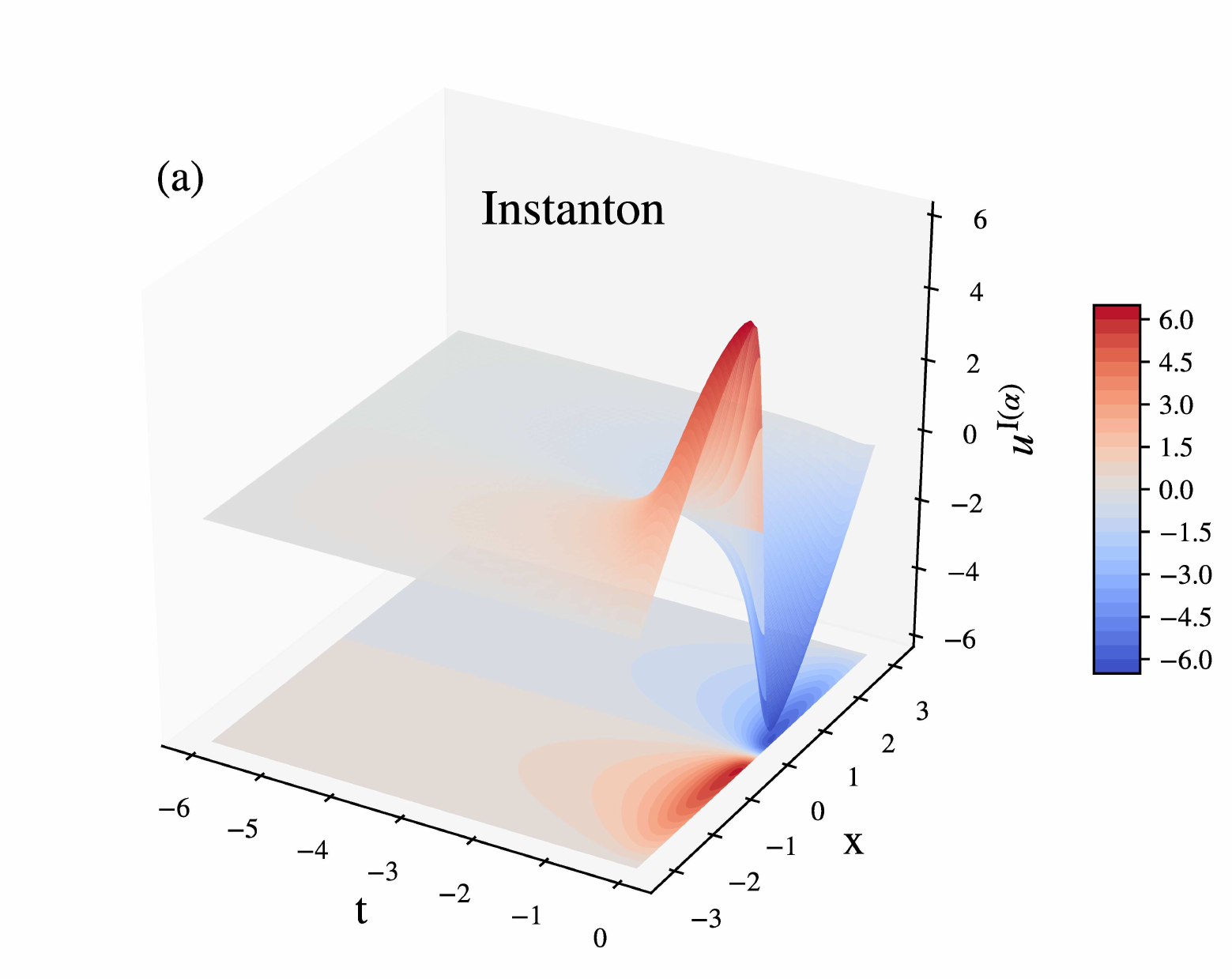}
    \includegraphics[width=0.5\textwidth]{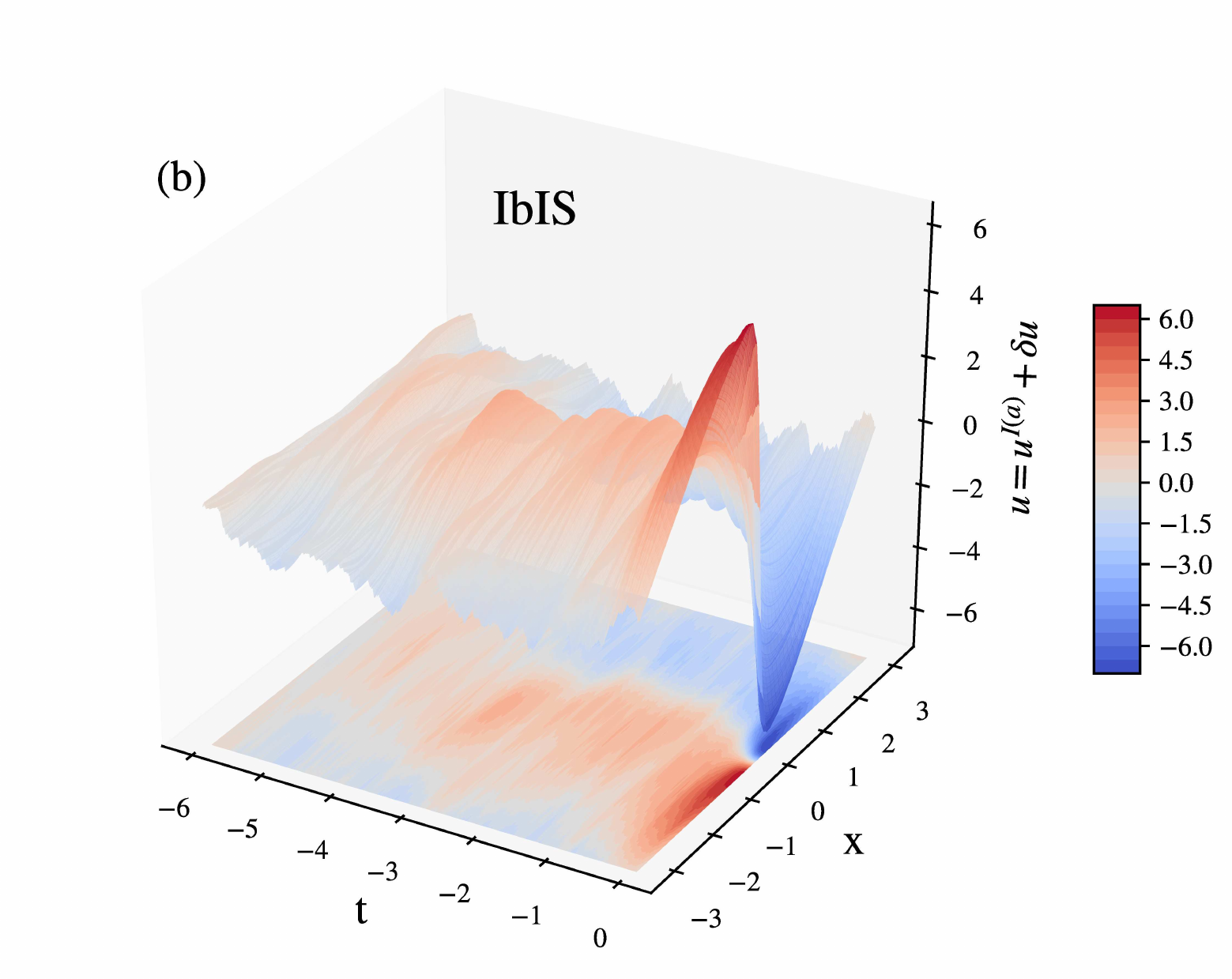}
	\caption{(a) Instanton solution $u^{I(a)}$ for $N_{x}=64$, $N_{t}=576$, $\nu=0.5$, $u^{I(a)}_x(0,0)=a=-31.8$.
	         (b) A realization $u(x,t)=u^{I(a)}(x,t) + \delta u(x,t)$, using the above instanton after solving Eq.~(\ref{eq:tilde_pde})}
	\label{fig:instanton}
\end{figure}

Results of an instanton configuration for a certain gradient
$u_x^{I(a)}(x=0,t=0) = a$, and a snapshot of a typical realization of
Eq.~(\ref{eq:tilde_pde}) $\delta u$ added with $u^{I(a)}$ are displayed in
Fig.~\ref{fig:instanton}.

\subsection{Generating the PDF}\label{subsec:gen_pdf}

In order to generate the PDF of velocity gradients, we define a set of
Lagrange multipliers $\mathcal{F}$ that implicitly define the set of
gradients $a$. For each of these Lagrange multipliers we complete the
following iteration scheme:
\begin{enumerate}
	\item Solve the instanton equations Eqs.~(\ref{eq:u_inst})-(\ref{eq:p_inst}) and save both the velocity field $u^{I(a)}$ and the auxiliary field $p^{I(a)}$ as well as the convolution $P^{I(a)}$ in space and time.
	\item  Calculate the instanton action $S^{I(a)}$ according to Eq.~(\ref{eq:inst_action}).
	\item For a chosen number of realizations $N$:
	\begin{enumerate}
		\item Calculate the fluctuations around the instanton as stated in
    section~(\ref{sec:reweighting}), whilst calculating the space integral
    \begin{equation}
			\frac{1}{2} \int dx\, p^{I(a)}_{x} (\delta u)^{2}
		\end{equation}
    at every time step, such that the sum over all time steps gives the space
    time integral from Eq.~(\ref{eq:delta_s})
		%
		that is required in order to calculate the reweighting factor
		\begin{equation}
			\Delta S^a = S^{I(a)} - \frac{1}{2} \int dt\, dx\, p^{I(a)}_{x} (\delta u)^{2} \	.
		\end{equation}
		\item Add the instanton and the fluctuation
    \begin{equation}
			u = u^{I(a)} + \delta u
		\end{equation}
    and subsequently calculate the gradient $u_{x}$ at the space-time point $(x,t) = (0,0)$.
		\item Create the histogram of $u_{x}$ around $a$, where the bin size corresponds
    to the spacing of the gradients $a$, and the current realization of $u_{x}(0,0)$ is
    weighted by the factor $e^{- \Delta S^a}$.
  \end{enumerate}
	\item Take the mean value of all the histograms to obtain the value of the PDF at $u_{x}=a$.
\end{enumerate}

This structure allows it to run the process in parallel, because each of the
levels in the iteration scheme is independent. First the iteration for each of
the Lagrange multipliers can be done in parallel as well as the subroutine for
each of the realizations of the fluctuations.

\subsection{Results}\label{subsec:results}

\begin{table*}[t!hbp]
	\begin{ruledtabular}
	\begin{tabular}{rrrrrrr}
		$\nu$ & \# meshpoints & \# timesteps & time interval & \# realizations & method & computing time (cpu hrs) \\
		0.5   &  64 &  576 & 6 & $1\times10^{9}$ & DNS & $1\times10^{3}$ \\
		0.5   &  64 &  576 & 6 & $6\times10^{5}$ & HMC & $1\times10^{3}$ \\
		0.5   &  64 &  576 & 6 & $170\times10^{5}$ & IbIS & 24 \\
		0.2   & 256 & 1152 & 4 & $2\times10^{8}$ & DNS & $2.7\times10^{3}$ \\
		0.2   & 256 & 1152 & 4 & $4\times10^{5}$ & HMC & $1.2\times10^{4}$ \\
		0.2   & 256 & 1152 & 4 & $180\times10^{5}$&  IbIS & 250
	\end{tabular}
	\end{ruledtabular}
	\caption{ \small  The parameters used for the numerical simulations. $\nu$ is the viscosity, \# meshpoints is the number of grid points $N_x$ in space, \# timesteps is the number of points $N_t$ in time, while time interval denotes the physical temporal length. By \# realizations we denote the number of produced space-time configurations. In the case of the IbIS method the notation $170\times10^{5}$ implies that we produced $10^{5}$ space-time configurations for each of the 170 instantons with a given $u_x(0,0)=a$. Accordingly for the HMC we produced $10^5$ configurations for each of the six values of $c_1$: (1.9, 1.6, 1.2, -1, -10, -20) for $\nu=0.5$, and of the four $c_1$: (0.9, 0.8, 0.6, 0.5) for $\nu=0.2$, which were finally combined. The computing time in cpu hours is the total budget required to produce the corresponding \# realizations, that were used in Fig.~\ref{fig:pdf}. Notice that the IbIS method is substantially cheaper than the HMC in providing similar quality of extreme events. }
	\label{table:parameters}
\end{table*}

\begin{figure}[h!tbp]
	\centering
	\includegraphics[width=0.5\textwidth]{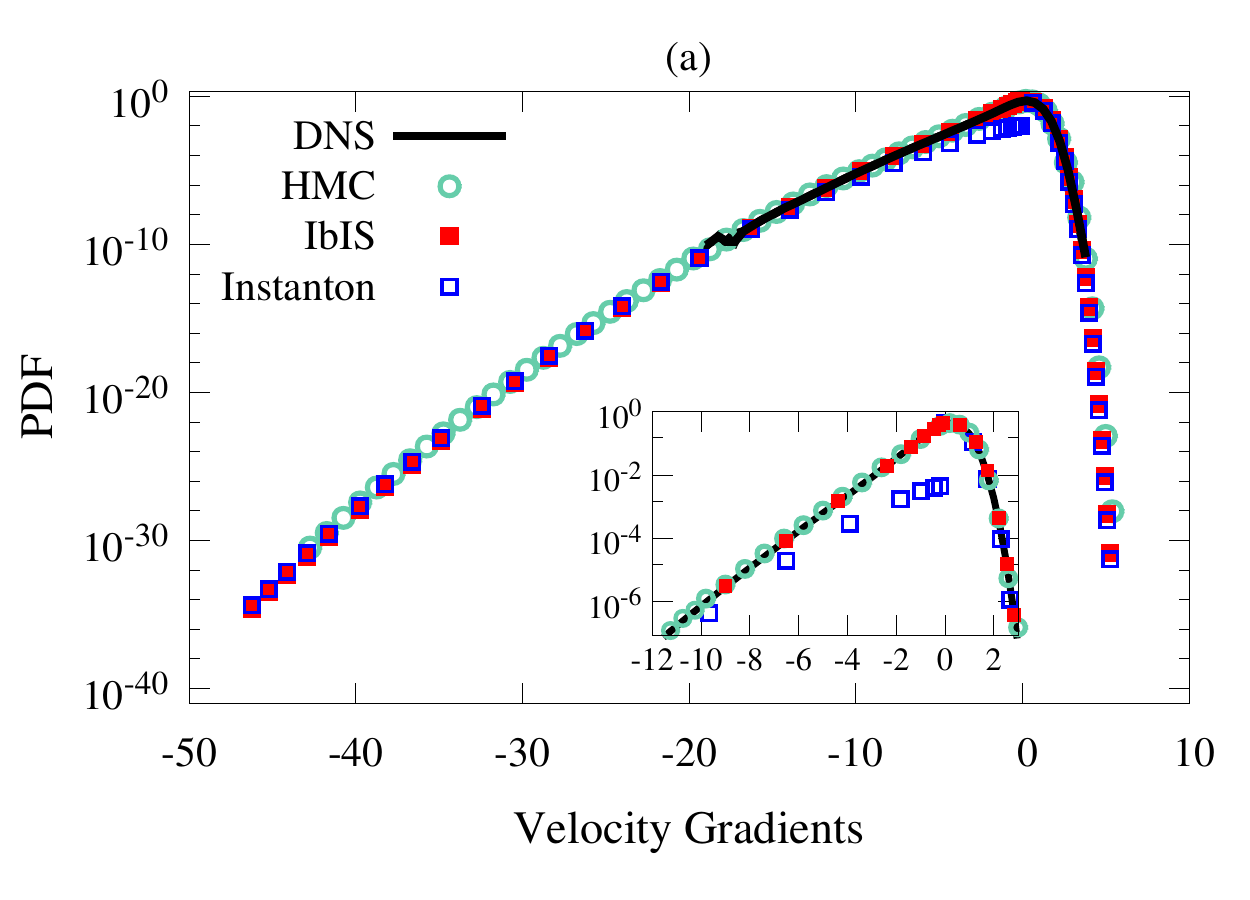}
    \includegraphics[width=0.5\textwidth]{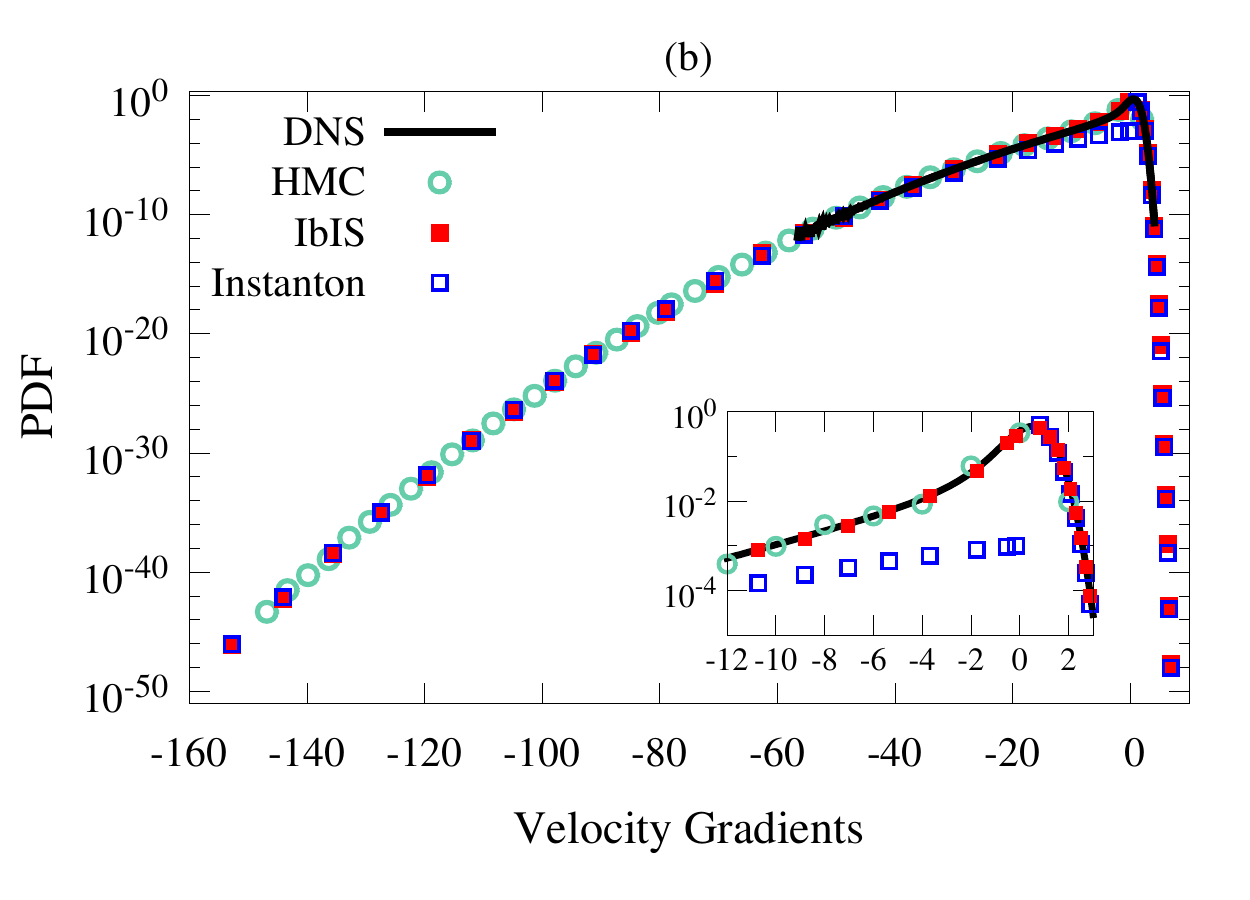}
	\caption{Velocity gradients PDF. Reweighted IbIS (red filled squares) and HMC (green open circles) versus the DNS (black line) and the instanton (open blue squares) (a) $N_{x}=64$, $N_{t}=576$, $\nu=0.5$, scanning for $a\in(-46,5)$. (b)  $N_{x}=256$, $N_{t}=1152$, $\nu=0.2$, scanning for $a\in(-160,10)$.}
	\label{fig:pdf}
\end{figure}

We performed two sets of simulations for two different Reynolds numbers
determined by the prescribed viscosities $\nu = 0.5$ and $\nu = 0.2$.  A set
consists of i) direct numerical simulations (DNS) of the Burgers equation, ii) a
hybrid Monte Carlo \cite{Margazoglou2018} (HMC) sampling of the path integral
and iii) our instanton based importance sampling (IbIS).

The hybrid Monte Carlo approach utilizes the action
$S_{\L}$~(\ref{eq:org_action}) that depends on the flow configuration and
constructs the measure as a weighted sum of all possible flow-realizations. Then
$S_{\L}$ together with an additional gradient maximization constraint
\begin{equation}
S' = S_{\L} + c_1 u_x(0,0)
\end{equation}
is sampled via the HMC algorithm \cite{Duane:1987de}, where the prefactor $c_1$ defines the strength of the constraint. The choice of the additional functional can, in principle, be arbitrary, here it is
specifically designed to systematically generate a large (positive, if $c_1<0$ or negative, if $c_1>0$) velocity
gradient at a specified space-time point, here at $(x,t)=(0,0)$. Hence, the system favors the sampling
of extreme and rare events in a similar spirit as an a posteriori filtering of
strong gradient events generated by a standard DNS\cite{Grafke:2013ska}.

To test its validity and capabilities,  the IbIS method is put in comparison
to the HMC and the DNS by measuring the PDF of the velocity gradients. Both
the IbIS and the HMC consider the reweighted statistics of the velocity gradient
measured only at $(x,t)=(0,0)$ (as explained in Sec.~\ref{subsec:gen_pdf} and in
\cite{Margazoglou2018}), while for the DNS we consider any site that belongs to
the stationary regime. In Figs.~\ref{fig:pdf}(a,b) we compare the cases of
$\nu=0.5$ and $\nu=0.2$, respectively. The inlet plot is an enlargement of the
central region of  the PDF, to strain that both reweighted PDFs of IbIS and HMC,
successfully reproduce it. On the other hand, as expected, the instanton
prediction for small negative velocity gradients is wrong and underestimates the
real PDF, as in this region the instanton approach is invalid, while in the case
of the right tails the instanton prediction is exact\cite{Gurarie:1995qc}.
Most importantly, the far left tail of the PDFs is reproduced identically
both from the IbIS and the HMC simulations. In addition, the PDFs approach the
instanton prediction for increasing $|u_x|$ as it is stated by large deviation
theory. This also constitutes a proof of concept of the IbIS method, as both
implementations are completely different and independent.

\begin{figure}[h!tbp]
	\centering
	\includegraphics[width=0.5\textwidth]{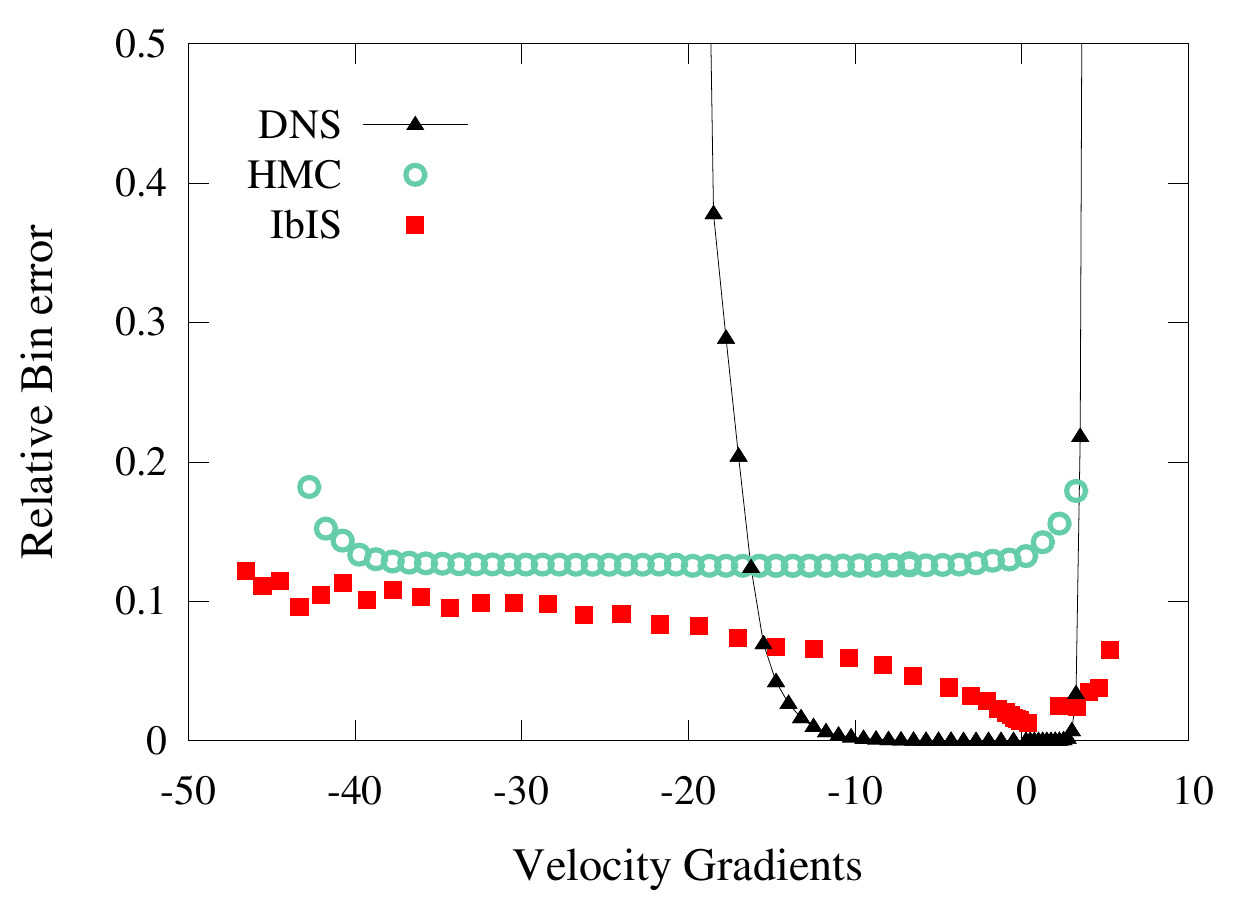}
	\caption{Relative bin errors of the velocity gradient PDFs of Fig.~\ref{fig:pdf}(a). The errors of the DNS quickly deviate, while the errors of the IbIS and the HMC stay small and under control.}
	\label{fig:pdf_error}
\end{figure}

In Fig.~\ref{fig:pdf_error} we plot the relative error of the bins in
Fig.~\ref{fig:pdf}(a). We notice that in the case of the DNS, the errors quickly
deviate for large velocity gradients due to the sparsity of the measurements.
This result is expected and underlines the need for rare-event algorithms in
turbulence. Contrary to the direct numerical simulations, both the IbIS and the
HMC provide sufficient statistics, resulting to constant and controllable small
relative errors over a substantially extended range of values of the PDF
of velocity gradient fluctuations ${\cal P}(u_x)$. In this respect, both the HMC and
IbIS strategy are of comparable quality to capture rare events in turbulence.

The difference between the HMC and the IbIS methods is captured in
Table~\ref{table:parameters}. This table does not only show all parameters used
in our simulations, but most importantly the run-time used for the different
simulations. First, we note the run-time for the DNS is about the size (or even
smaller) than the run-time used in the HMC simulations. However, we stress that
the DNS is only capable to capture a tiny fraction of the PDF. The remarkable
effectiveness of the IbIS compared to the HMC method can be deduced from the
ratio of their computing times. Here for the parameters used in our test cases,
the IbIS method turns out to be two orders of magnitudes faster than the HMC
approach, a ratio that is even expected to increase for higher Reynolds numbers.

%

\section{Conclusions and Outlook}

In this paper we have presented a new method based on instanton importance sampling to calculate the probability distribution function of velocity gradients in Burgers equation for both typical and extremely intense events. By sampling fluctuations of the SPDE obtained on the background of one given  instanton with a fixed (negative) intense velocity gradient in a certain $(x,t)$ position,  we explore the fluctuations around that specific flow configuration. At varying the reference gradient, and with a suitable reweighting protocol, we are able to reconstruct the whole PDF. The method is fully general and can  --in principle-- be applied to any SPDE.  To be successful, a necessary condition is that a large deviation principle is
applicable that guarantees the availability
of unique instanton solutions. The IbIS method will work most efficiently when the PDF obtained solely from the instanton prediction is  not to far from the true PDF. We also compared this new method with  a  Hybrid Monte Carlo approach \cite{Margazoglou2018}  which does not rely on these assumptions and thus is applicable to a larger variety of physical problems. Concerning the Burgers case, IbIS is orders of magnitudes faster then the HMC. Both methods are much better than standard pseudo-spectral algorithms  which are unable to focus on extreme-rare events.  With the IbIS method it might be possible to calculate the scaling of the algebraic power law prefactors, which characterizes the  inviscid limit of Burgers equations
\cite{e-khanin-mazel-sinai:1997,e-eijnden:2000a} without using
Lagrangian particle and shock tracking methods\cite{bec:2001}.

\begin{acknowledgments}
J.F. acknowledges funding from the Humboldt foundation within a Feodor-Lynen fellowship.
L.B. acknowledges funding from the European Research Council under the European Union's Seventh Framework Programme, ERC Grant Agreement No. 339032.
G.M. kindly acknowledges funding from the European Union's Horizon 2020 research and innovation Programme under the Marie Sk\l{}odowska-Curie grant agreement No 642069 (European Joint Doctorate Programme "HPC-LEAP").
\end{acknowledgments}

\bibliography{references}

\end{document}